# Retention and Deformation of the Blue Phases in Liquid Crystalline Elastomers

Kyle R. Schlafmann[1], Timothy J. White[1,2,*]

[1]Department of Chemical and Biological Engineering, University of Colorado – Boulder

[2]Department of Materials Science and Engineering, University of Colorado – Boulder

*Correspondence to TJW - Timothy.J.White@colorado.edu

**Abstract**

The blue phases are observed in highly chiral liquid crystalline compositions that nascently organize into a three-dimensional, crystalline nanostructure. The periodicity of the unit cell lattice parameters is on the order of the wavelength of visible light and accordingly, the blue phases exhibit a selective reflection as a photonic crystal. Here, we detail the synthesis of liquid crystalline elastomers (LCEs) that retain blue phase I, blue phase II, and blue phase III. The mechanical properties and deformation of LCEs retaining the blue phases are contrasted to the cholesteric phase in fully solid elastomers with glass transition temperatures below room temperature. Mechanical deformation and chemical swelling of the lightly crosslinked polymer networks induces lattice asymmetry in the blue phase LCE evident in the tuning of the selective reflection. The lattice periodicity of the blue phase LCE is minimally affected by temperature. The oblique lattice planes of the blue phase LCEs tilt and red-shift in response to mechanical deformation. The retention of the blue phases in fully solid, elastomeric films could enable new functional implementations in photonics, sensing, and energy applications.

**Introduction**

Liquid crystalline elastomers (LCEs) are compelling responsive materials[1]. Recently, they have drawn considerable research interest for their utility as machine-like mechanical actuators[2] as well as the spatial programmability of their material properties[3]. Analogous to small-molecule liquid crystalline systems, which have been prevalent in display applications for some time, LCEs also are relevant to a variety of optical and photonic applications including solid-state optical sensing[4], tunable diffraction[5], and adaptive lasing[6].

Here we detail an approach to prepare fully solid photonic materials that are robust, yet elastic and capable of dynamic reconfiguration. The fluidic nature of small molecule liquid crystalline systems enables the response time and performance of liquid crystal displays[7]. In certain functional implementations, the realization of dynamic optical response in fully solid films is particularly desirable. Further, as evident in biological systems, dynamic reconfiguration of solid skins naturally induces multifunctionality even under ambient conditions[8]. Ranging from regulated structural color and diffuse reflectance in cephalopods[9,10] to glare-reducing corneal nipple arrays on nocturnal insect eyes[11,12] and photonic crystalline structures responsible for coloring in butterfly wing scales[13,14], the natural environment provides countless examples of dynamic light control. Many synthetic material systems have incorporated bio-inspired mechanisms to imitate their photonic properties for technological applications[15].

Multi-dimensional photonic crystals such as those observed in the structural coloration in certain animals are of particular interest for optical applications such as non-linear optics and waveguides[16]. These unique periodic nanostructures exhibit photonic band gaps in multiple propagation directions. Three-dimensional photonic crystals are nano-scale emulators of crystal structures typically observed in atomic lattices. Nanostructures such as those arising from inverse opals or lithographic deposition exhibit three-dimensional photonic crystallinity, however the intensive fabrication of such materials is complex and costly[17]. The optics and photonics community has increasingly looked towards self-assembly and liquid crystals as light-manipulating media. Liquid crystalline materials can be formulated to adopt a catalogue of anisotropic phases. Chiral liquid crystalline phases in particular exhibit photonic band gaps due to their periodic anisotropy[18].

The liquid crystalline blue phases are a subset of liquid crystalline phases in which calamitic mesogens align in a double-twist morphology associated with large concentrations of chiral species. The blue phases are frustrated phases and the nanostructure of these materials are a combination of defect packing and double-twist molecular arrangement of cylinders on the order of 100nm in diameter[19]. These cylinders stack upon one another to generate cubic lattices stabilized by disclinations. Blue phase I (BPI) organizes as a body-centered cubic (bcc) lattice, blue phase II (BPII) achieves a simple cubic (sc) lattice, and blue phase III (also called the "blue fog") is largely amorphous. In this way, BPI and BPII are three-dimensional photonic crystals that are selectively reflective to visible light[19].

The cubic blue phases have been extensively studied in non-crosslinked low molar mass systems[20–27]. The primary motivation for these studies is the potential for fast electro-optical switching properties in devices that do not need alignment layers or precise thickness control[28]. They have also been retained in free-standing glassy polymers upon crosslinking within the thermotropic phase window[29–34]. These polymers exhibit largely static photonic properties due to their high crosslink density, however recently the selective reflections of these materials have been sensitized to humidity and pH through supramolecular bonds within the network[33] and to small amounts of mechanical deformation[34]. Polymer-stabilized blue phase gels (32 wt.% polymerizable media) exhibit electro-optical response, along with limited photonic sensitivity to mechanical deformation[35]. However, the liquid crystalline blue phases have not been retained in truly elastomeric polymer networks with a glass transition below room temperature ($T_g < 20°C$).

We have reported a straightforward and scalable approach to prepare well-aligned cholesteric liquid crystal elastomers (CLCEs) in the planar orientation composed of main chain chiral mesogens[36]. Here, enabled by a similar reaction chemistry, we prepare liquid crystalline elastomers that retain the blue phases upon photopolymerization. We characterize changes in the lattice spacing to mechanical deformation, heat, and chemical exposure.

**Results and Discussion**

Historically, liquid crystalline elastomers (LCEs) have been prepared by two-stage polymerization reactions in which alignment is enforced in the second stage by mechanical force. The preparation of hierarchical liquid crystalline phases, such as the cholesteric or blue phases, are not readily amenable to these processes. The helicoidal nanostructure of the cholesteric phase has been retained in LCEs by mechanical alignment coupled with anisotropic deswelling through either centrifugation[37–39] and/or anisotropic deswelling[40]. However, the intricate, 3-dimensional nanostructure of the blue phases have not been retained in LCEs by these methods. A recent report[36] details the development of a materials chemistry to prepare LCEs that originate from the desired chiral phase that are conducive to surface-enforced alignment and prepared from a one-step reaction. This chemistry was utilized to prepare and retain the cholesteric phase in LCEs of high optical quality (low haze) and considerable thermochromism.

The blue phases are observed in formulations that have a large concentration of chiral species. The liquid crystalline molecules organize in cubic, nanostructured phases. The periodicity of the blue phases is defined by crystalline packing within unit cells. The blue phases are typically observed in very narrow temperature ranges, which can complicate the retention of these phases in polymer networks. Here, we report the synthesis of LCEs that retain blue phase I (BPI), blue phase II (BPII), and blue phase III (BPIII). The composition is based on a mixture of the liquid crystalline diacrylate C6M and chiral liquid crystalline diacrylate SLO4151 (**Figure 1a**). The non-reactive chiral dopant R811 is added at 15 wt% to increase the concentration of chiral species. Notably, the addition of R811 both broadens the temperature range of the blue phases before polymerization (**Figure S1**) and enables the formation of

BPII and BPIII in these mixtures. We employ a one-step free-radical photopolymerization reaction between the acrylate monomers and 14.2 wt% of the dithiol, BDMT. BDMT can react as both a chain extender and chain transfer agent[41]. Polymerization is photoinitiated with Omnirad 819. A representative DSC trace of the LCE is illustrated in **Figure 1b** (individual traces found in **Figure S2**). All the LCE detailed in this examination have below-room temperature glass transition temperatures ($T_g$) of approximately 15°C. The acrylate-thiol chain transfer reactions suppress the crosslink density of the acrylate homopolymerization by producing a highly branched polymer network architecture illustrated in the inset to **Figure 1b**[41].

The formulation to prepare LCEs with 15wt% R811, 14.2 wt% BDMT, liquid crystalline monomers, and photoinitiator exhibits five liquid crystalline phases before polymerization. Accordingly, this study is uniquely able to prepare LCEs in the nematic (polydomain), cholesteric, BPI, BPII, and BPIII phases simply by varying the polymerization temperature. Upon polymerization, the LCEs retain the characteristic birefringent textures of these phases when imaged with polarized optical microscopy (POM, **Figure 1c**). To confirm the retention of cubic periodicity in LCEs, we undertook so-called Kossel diffractive measurements[24]. As expected, the cholesteric, BPIII, and nematic phases do not exhibit a diffraction pattern (**Figure 1d**). However, the cubic orientation of BPI and BPII are evident (**Figure 1d**). The photographs of the LCE retaining the cholesteric, BPI, and BPII illustrate the retention of the selective reflection associated with these phases. A hazy blue texture indicates the amorphous BPIII phase, which is not cubic (evident in **Figure 1d**) and does not exhibit a coherent reflection. Extraction of the non-reactive chiral dopant from the LCE results in a slight blue-shift in reflection color, evident in the images of the free-standing LCE films (**Figure 1e**) when compared to the POM images in Figure 1c. (see also **Figure S3**). Gel fraction measurements confirm complete incorporation of the reactive components and removal of the non-reactive chiral dopant.

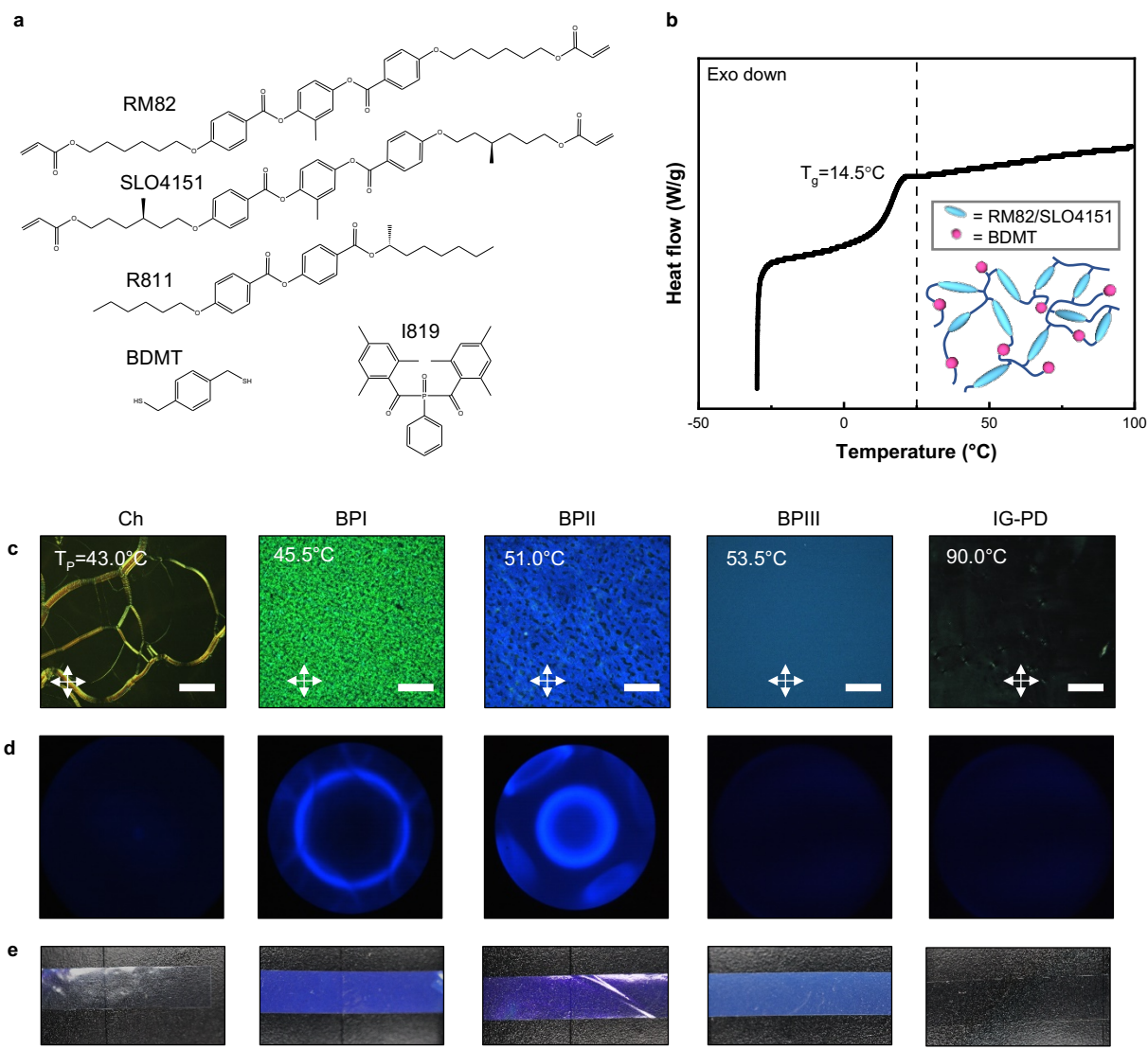

**Figure 1 | Blue Phase LCE formulation and phase retention**. **a**) Chemical structures of monomers. **b**) DSC thermogram illustrating glass transition of polymer network, independent of the phase retained. Inset) the LCE forms a hyper-branched polymer network. **c**) Polarized optical micrograph textures (in reflection mode) of the LCE. Scale bars are 100 μm. Images are captured at room temperature after each phase is retained via photopolymerized at the polymerization temperature ($T_P$). From left to right: cholesteric phase (Ch), blue phase I (BPI), blue phase II (BPII), blue phase III (BPIII), and isotropic genesis polydomain (IG-PD). **d**) Kossel diagram (440 nm) of free-standing elastomers retaining the stated phase confirming cubic structures for blue phase I (body-centered cubic [110]) and blue phase II (simple cubic [100]). **e**) Photographs of LCE retaining the either Ch, BPI, BPII, BPIII, or IG-PD phase, unpolarized illumination.

Informed and enabled by the retention of the blue phases in fully solid and elastomeric LCE, we focus the remainder of this examination on LCE retaining the BPI and BPII phases. The selective reflection of unperturbed BPI and BPII LCE, visualized in the photographs in **Figure 1e**, were quantified with UV-vis spectroscopy (**Figure 2a**). These spectra are contrasted to the LCE retaining the cholesteric phase, which due to the large concentration of chiral monomer and chiral dopant used to prepare these LCE, exhibits a selective reflection below 350 nm (appearing transparent in **Figure 1e**). UV-vis spectra for BPIII LCE and nematic (polydomain) are shown in **Figure S4**. The mechanical deformation of LCE retaining BPI, BPII, and the cholesteric phase are evident in **Figure 2b**. The stress-strain response of BPI LCE and BPII LCE are non-linear between 25% and 70% strain before transitioning to a strain hardening region. Comparatively, the deformation of the cholesteric LCE (CLCE) retained from the identical composition exhibits strain hardening around 50% strain. The deformation of BPI, BPII, and BPIII LCE resembles the soft elasticity typically observed in isotropic genesis polydomain LCE (**Figure S5**). We attribute this similarity to the double-twist cubic nanostructure which is macroscopically isotropic in three-dimensions. Comparatively, while the director in CLCE introduces effective isotropic mechanical properties in the x-y plane, the planar organization of these materials does differentiate the mechanical properties in the z axis (through the thickness). Thus, we conclude the phase of the LCE is the primary differentiator in stress-strain behavior reported here.

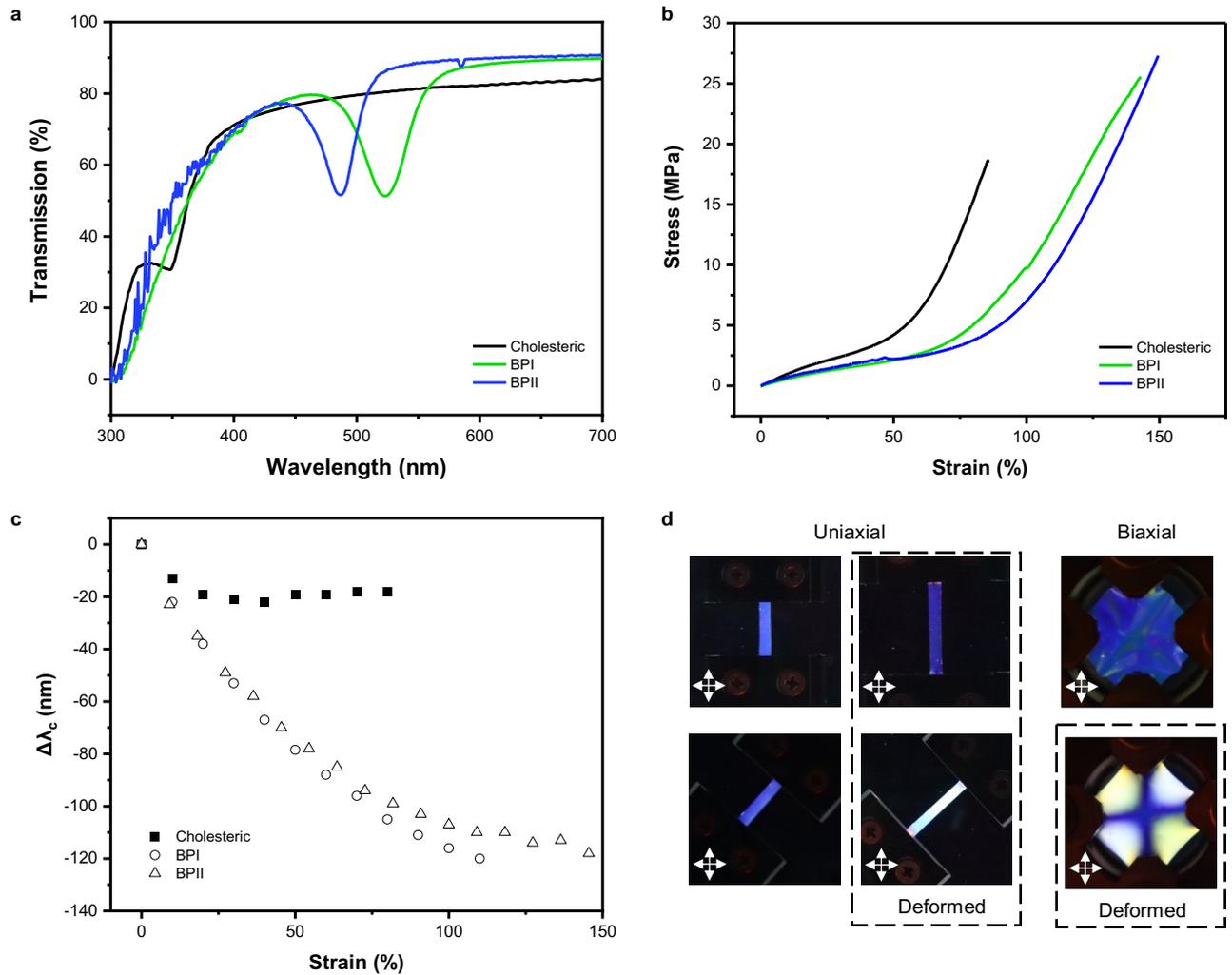

**Figure 2 | Optical properties and effects due to mechanical deformation**. **a**) Transmission spectra of elastomers retaining the cholesteric phase, blue phase I (BPI), and blue phase II (BPII) showing selective reflection bands as measured through incident light to the surface of the elastomer. **b**) Stress-strain behavior of cholesteric, blue phase I, and blue phase II elastomers. **c**) change in selective reflection of elastomers retaining the cholesteric phase, blue phase I (BPI) and blue phase II (BPII) as a function of uniaxial strain. **d**) Birefringence patterns induced by uniaxial or biaxial mechanical deformation of blue phase I elastomer. Observed as photographs through crossed polarizers.

The mechano-optical response of the deformation of BPI and BPII was examined by UV-vis spectroscopy during an applied load. The selective reflection ($\lambda_{hkl}$) of the cubic blue phases (BPI and BPII) is defined by:

$$\lambda_{hkl} = \frac{2\bar{n}a}{\sqrt{h^2+k^2+l^2}} \qquad (1)$$

where $\bar{n}$ is the average refractive index, $a$ is the lattice constant, and $h, k, l$ are the Miller indices. Uniaxial deformation lengthens the material in the loading axis while reducing the width of the film and more importantly, the material thickness. Again, the optical properties of BPI and BPII LCE are derived from periodic cubic nanostructures. The directional bias introduced by mechanical load affects the lattice spacing within the BPI and BPII which cause the LCE retaining these phases to undergo a sizable blue shift in the primary reflection. The magnitude of this shift (120 nm) is greater than that observed for the CLCE polymerized from the same formulation. The reflection of the BPI LCE is associated with the [1,1,0] lattice plane while the reflection of the BPII LCE is associated with the [1,0,0] lattice plane. The blue shift in the deformation of a BPI LCE and BPII LCE are visually evident in **Figure 2c**. Note, due to the position of the [1,0,0] reflection of BPII LCE, deformation nearly immediately shifts the reflection into the ultraviolet (UV). Analogous to BPII phases in low-molar mass mixtures, the BPII LCE maintains a circularly polarized reflection that match the handedness of the chiral origins. To deformation of 50% strain, the reflection of both BPI and BPII LCE becomes unpolarized (**Figure S6**). A conversion from polarized to unpolarized reflection has been predicted[42] and observed[43] in CLCE, owing to asymmetry in the period distribution of refractive index associated with thickness-dependent (orientation-dependent) reorientation to the strain axis,. Biaxial strain of BPLCE results in an alternating pattern of yellow and blue regions which we believe may be related to a distinctive mechanical accommodation of the double-twist cylinder mesogen geometry.

The influence of uniaxial deformation of BPI and BPII LCE to introducing cubic asymmetry is further elucidated by Kossel imaging. In this method, monochromatic light (440 nm) is incident upon the blue phase LCE through a high numerical aperture objective (**Figure 3a**). The rear focal plane is observed using a Bertrand lens. The incident light exhibits a characteristic diffraction pattern as it passes through the cubic lattice of the blue phase LCE. The two-dimensional projection of the reciprocal lattice space is captured (**Figure 3b-c**).

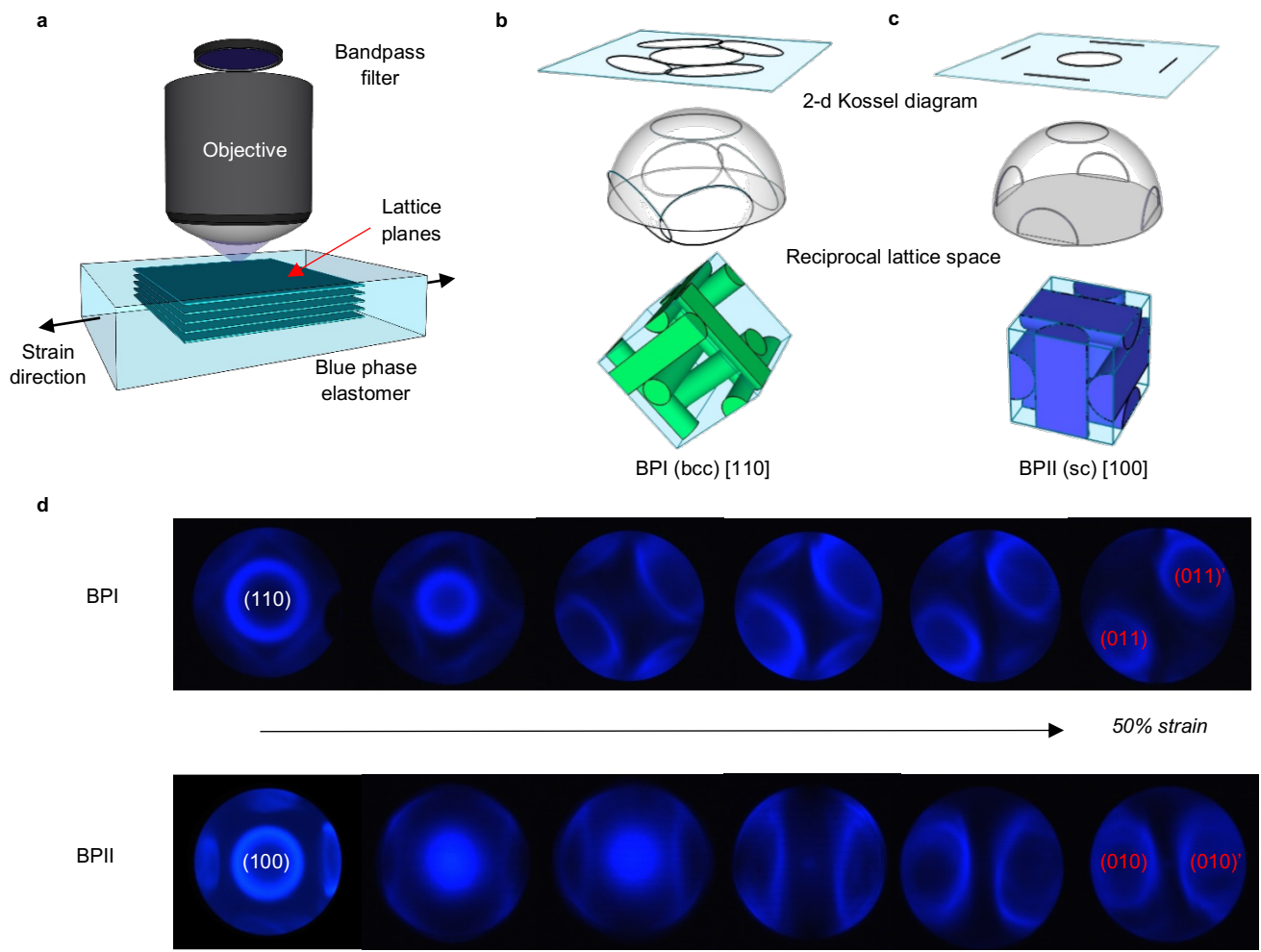

**Figure 3 | Blue phase LCE lattice deformation (Kossel diagram analysis). a**) Kossel diagram capture apparatus. A polarized optical microscope was employed with Bertrand lens, a 100x objective lens and 440 nm bandpass filter. **b**) (top) Theorized Kossel diagram of [110] oriented blue phase I. (middle) Reciprocal lattice space of [110] oriented blue phase I. (bottom) Unit cell of [110] oriented blue phase I. **c)** (top) Theorized Kossel diagram of [100] oriented blue phase II. (middle) Reciprocal lattice space of [100] oriented blue phase II. (bottom) Unit cell of [100] oriented blue phase II. **d)** Kossel diagrams of blue phase I (BPI) and blue phase II (BPII) elastomers as a function of (horizontal) uniaxial strain, showing unit cell reconfiguration.

Evident in **Figure 3d**, in the undeformed state, the BPI LCE (bcc) maintains [110] lattice alignment in the viewing direction with Kossel lines corresponding to the adjacent [011] and [101] planes at approximately the expected angle of 60° from the viewing plane. As the BPI LCE is subjected to uniaxial load, the circle corresponding to the [110] plane decreases in diameter. The decrease indicates that the lattice spacing of the [110] plane is decreasing, which affects the periodicity of the nanostructure.

Simultaneously the arc Kossel lines corresponding to the [011] planes become more circular with strain, while the Kossel lines corresponding to the [101] planes depart the field of view. We believe this confirms the [011] lattice planes are tilting toward the film surfaces during deformation. Additionally, the periodicity of this plane does not decrease by the same magnitude as the vertically oriented [110] plane. This indicates that the uniaxial load induces lattice asymmetry.

The BPII LCE (sc) maintains [100] lattice alignment in the viewing direction with Kossel lines corresponding to the [010] and [001] planes observed at the expected angle of 90°. Similarly, the Kossel diagrams of BPII LCE during uniaxial deformation confirm that the lattice spacing becomes asymmetric to deformation. The simple cubic [100] periodicity decreases with strain. In an analogous fashion to the deformation-induced lattice tilting of BPII LCE, the [010] or [001] (whichever is oriented along the strain-direction) tilts towards the film surface and maintains a periodicity much larger than that of the normal incidence [100] in the deformed state.

Again, the liquid crystalline blue phases are cubic structures; their selective reflection originates from the periodicity of that cubic lattice. In that sense, each set of lattice planes act as a Bragg reflector which blue-shifts $\propto \cos(\theta)$ as the viewing angle increases. To further explore the spectral effect of mechanical deformation of the lattice, **Figure 4** measures the angular dependent optical properties of a BPII LCE during deformation. Evident in **Figure 4a**, at 0% strain the BPII LCE exhibits blue-shifting when rotated from orthogonal to the optical probe to angled 45°. At normal incidence (e.g., 0°), the reflection is primarily associated with the [100] plane. However, when probing the BPII LCE at 45° to the film surface, both the [100] and [010]/[001] (whichever is aligned along the long axis of the film) are observable. Both the [100] and [010]/[001] are each oriented at 90° angles to one another in a simple cubic lattice. Therefore, we would expect an undistorted BPII LCE to have a single, blue-shifted reflection band when measured at 45°, because the $\cos(\theta)$ dependence is equivalent. Again, the extraction of the non-reactive chiral dopant after preparation of these films may impart some residual stress that could be the source of the lattice tilt that may be the origin of the small reflection apparent at

460 nm. The photographs of the unstrained BPII LCE in **Figure 4b** demonstrate a slight blue-shift from blue to purple between the normal incidence and 45° sample angle.

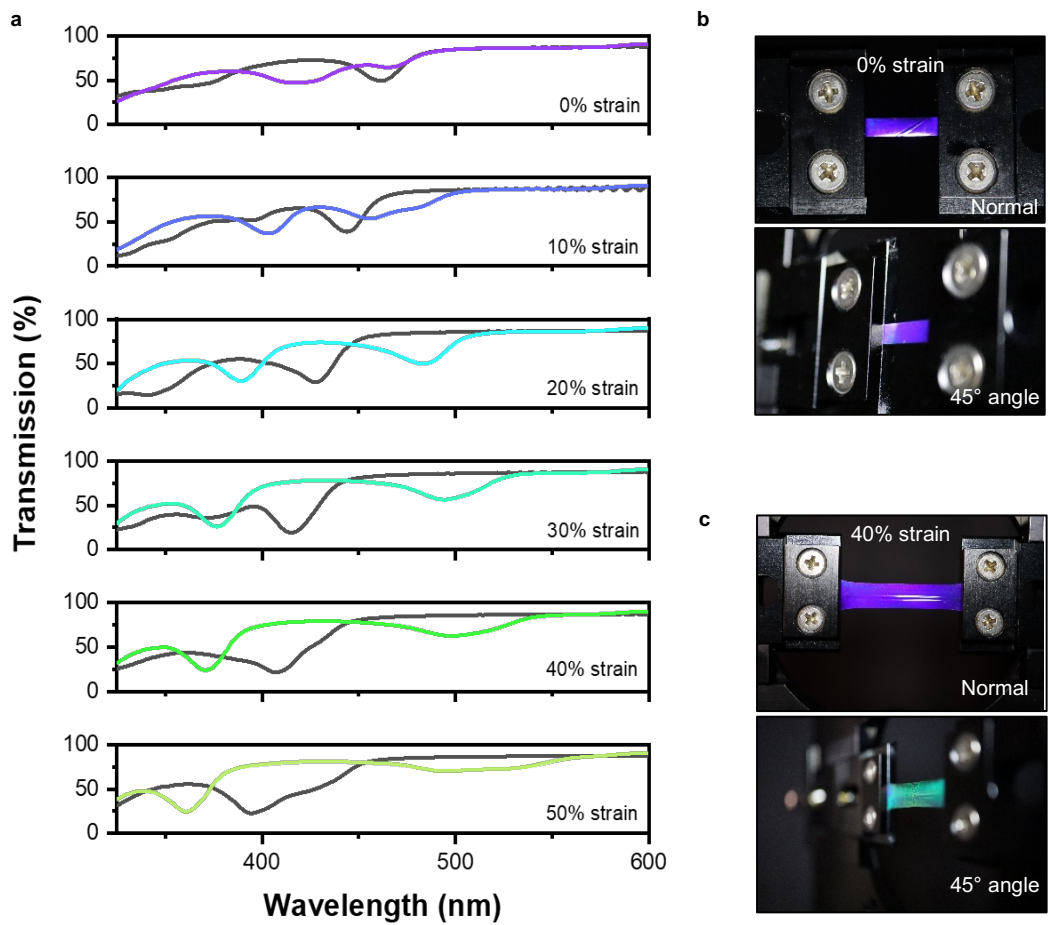

**Figure 4 | Optical confirmation of BPII LCE lattice asymmetry with deformation. a**) Transmission spectra of a BPII LCE with (i) the surface of the film at normal incidence to the optical probe [black] and (ii) the film angled 45° to the optical probe [color]. **b**) Photos of unstrained BPII LCE illuminated from camera direction. (top) Sample at normal incidence (bottom) 45° angle. **c**) Photos of 40% strained BPII LCE illuminated from camera direction. (top) Sample at normal incidence (bottom) 45° angle.

As the sample is uniaxially deformed, the reflection associated with the [100] lattice plane at normal incidence begins to blue-shift (as shown in **Figure 2b**) due to the decreasing film thickness. The [100] reflection notch is blue-shifted further when the probe-sample-detector is at an acute angle, due to the Bragg condition. However, the [010]/[001] reflection at 45° BPII LCE begins to red-shift as the sample is strained. We attribute this to a combination of lattice tilt towards the surface of the film (as shown in **Figure 3d**) and an increase in [010]/[001] periodicity. Our underlying assumption is that the total number of unit cells is unchanged in the crosslinked polymer network and thus, the redistribution of volume of the LCE is accommodated by an asymmetry within the unit cells. The lattice constants of the unit cell become shorter in the [100] plane direction and longer in the strain-axis lattice plane direction. This effect is observable visually in the photographs in **Figure 4c**. Of particular note, the red-shifted reflection of the [010]/[001] BPII LCE at 40% strain is apparent when imaged at an acute angle.

The alignment of blue phase materials has drawn considerable interest as of late due to the significant improvement in coherent reflective properties of the material. Without an alignment layer, BPI or BPII unit cells can orient randomly to the material surface and the phase is retained in a polycrystalline state. Due to natural differences in refractive indices, the polycrystalline blue phases can exhibit significant haze. Recent literature reports have documented the critical contribution of surface interaction to enabling alignment of large monocrystalline blue phase domains[21]. Here, we demonstrate that the domain size and optical quality of the reflection of LCE retaining BPI or BPII can also be affected by variations in surface anchoring. Here, utilizing a commercially available photoalignment mixture PAAD-22 (BEAM Co.), an alignment cell was prepared and selectively irradiated with a 405 nm laser (**Figure S7**). As evident in **Figure 5a**, the irradiated (patterned) region exhibits a bright green reflection while the masked (un-patterned) region appears hazier and exhibits a faint green reflection. The resolution and effectiveness of the photoalignment mixture is further explored with polarized optical microscopy. The stark difference in platelet size and degree of orientation is evident in **Figure 5b**. While the aligned (irradiated) region is still polycrystalline (as evident by the dark regions in which the platelets reflect UV light), the degree of orientation is considerable. This discrepancy in reflection coherence is further exemplified in the transmission spectra and corresponding Kossel diffraction diagrams in **Figure 5c**. While the photoaligned region of the BPI LCE exhibits a single, selective reflection at 520nm, the un-patterned surface exhibits low transmission in the 350-500 nm range associated with random orientation of BPI unit cells. The hazy Kossel diffraction image is further indication of the dispersity of crystalline orientation in the un-patterned regions.

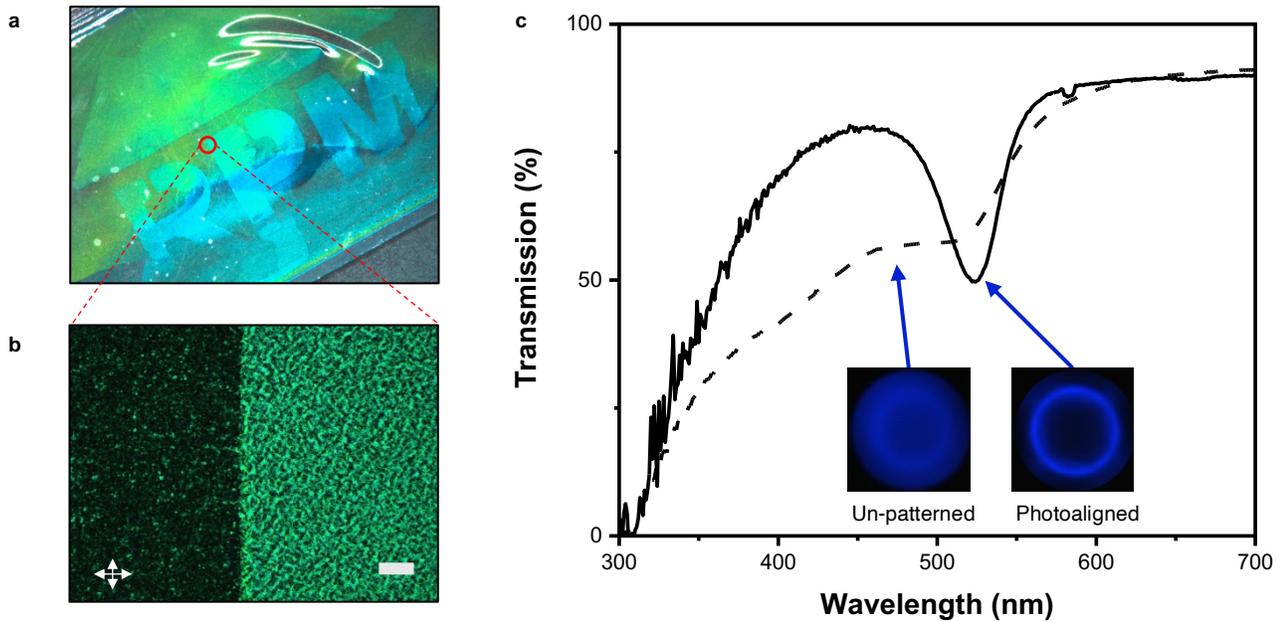

**Figure 5 | Patternable lattice alignment. a**) Demonstration of photopatterned blue phase I lattice alignment. To prepare, a photomask was employed to selectively irradiate the photoalignment cell with 405 nm light. **b**) Polarized optical micrograph texture imaged at the edge of the photopatterned region (scale bar 100 μm). **c**) Transmission spectra of BPI LCE un-patterned and photo-aligned regions. Kossel diagrams qualitatively assess lattice homogeneity.

Upon heating, the BPI and BPII LCE exhibit minimal change in reflection (**Figure 6a**). This is distinctive from an LCE that retain the cholesteric phase, that exhibits upwards of 200 nm shifts in selective reflection to changes in temperature[36]. As detailed above relating to the mechanical response, the effective macroscopic isotropy of BPI and BPII suppress the magnitude of the relative change in periodicity associated with the average thermal expansion coefficient of the polymer. Comparatively, LCEs retaining the CLC phase are isotropic in the x-y axes but anisotropic in the thickness, which results in a directional mechanical response to the thermotropic disruption of order.

Comparatively, swelling the BPI and BPII LCE with an organic solvent can yield sizable changes in coloration. The red-shift evident in the selective reflection of the POM micrographs in **Figure 6b** are associated with a 3-dimensional, symmetric increase in lattice spacing. A similar effect is observed in the solvent exposure of the initially UV-reflecting LCE retaining the CLC phase. The magnitude of the expansion in periodicity is solvent dependent. The interaction between the solvent and the chemical environment within the elastomer matrix can be defined by solubility parameters, which quantify compatibility of dispersion forces, polarity, and hydrogen bonding[44]. Swelling with toluene (nonpolar) resulted in comparatively larger uptake that the polar solvent acetone, which is qualitatively evident in the magnitude of the shift in reflection color. Exposure to dichloromethane resulted in even greater shifts into the near infrared spectrum (see Supplemental Video).

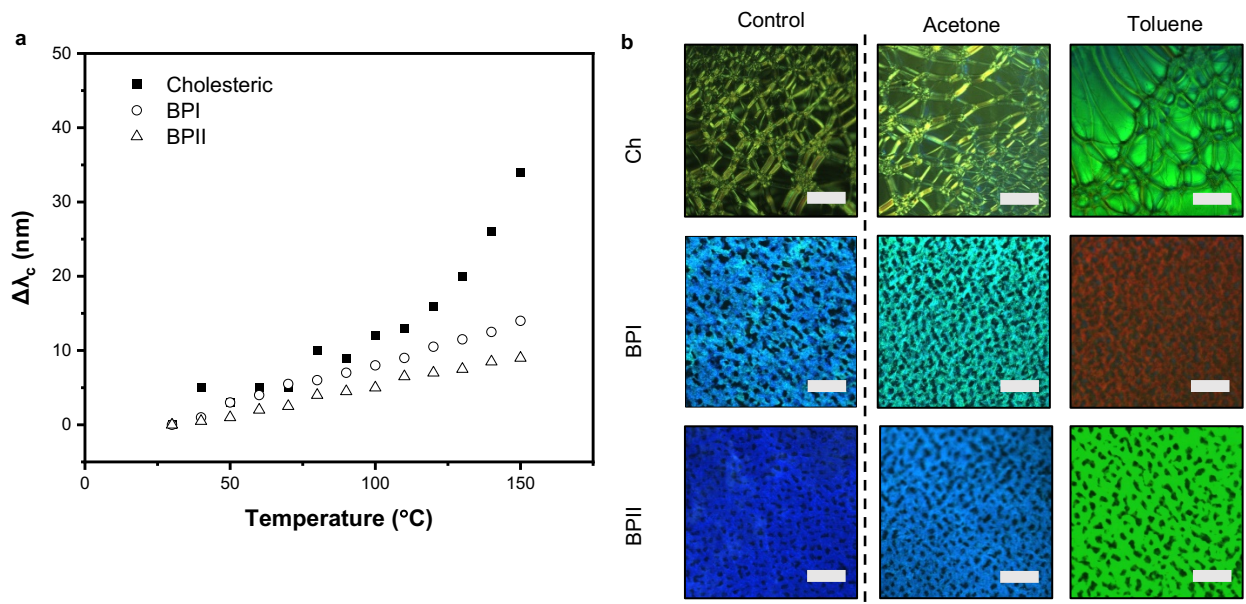

**Figure 6 | Optical response to chemical swelling and heat. a**) Selective reflection band wavelength of CLCE, BPI LCE, and BPII LCE as the samples are heated from room temperature to 150°C. **b**) Polarized optical micrograph textures observed in reflection mode of CLCE, BPI LCE, and BPII LCE (scale bar 100 μm).

The branched, lightly crosslinked polymer network prepared from the liquid crystalline monomer mixture reported here can retain the three-dimensional, nanostructured architecture associated with either the body-centered cubic BPI or the simple cubic BPII. Mechanical, thermal, and chemical stimuli uniquely affect the lattice constants and the associated optical properties of these free standing and fully solid optical materials.  Whether the stimuli-exposure introduces symmetric (heat, swelling) deformation or asymmetric (mechanical force) deformation of the lattice of the blue phases, in all cases the distortion is reversible. This novel class of lightweight, fully solid, and deformable optical materials have many compelling properties of potential benefit to functional use in non-linear optics, lasing, spectral imaging, and sensing.

## Methods

*Materials synthesis*

The achiral diacrylate C6M (Wilshire Technologies), the right-handed chiral diacrylate monomer SLO4151 (Alpha Micron), right-handed chiral dopant R811 (EMD Chemicals) were used as received. Blue phase LCE mixtures were prepared by adding 51 wt% RM82, 19 wt% SLO4151, 15 wt% R811, 14 wt% BDMT (benzenedimethanethiol, Sigma Aldrich), and 0.5 wt% Omnirad 819 (IGM Resins).

Alignment cells were made by spincoating two Corning EXG glass slides with 22 μL of photoalignment dye PAAD-22 (Beam Co.) diluted (2 to 1 by volume) with DMF. The coated glass slides were dried at 100°C for 30 minutes to drive off any residual solvent. Once dried, the coated sides of two glass slides were adhered together with an epoxy adhesive mixed with 20 μm glass spacers. Photoalignment was achieved by irradiating the alignment cell with a linearly polarized 405 nm laser at an intensity of 10 mW/cm$^2$ for 10 minutes. Spatial variation in alignment utilized a photomask.

To prepare the liquid crystal elastomers retaining the various phases detailed hereto, the monomer mixture was melted to an isotropic phase at 90°C and filled via capillary action into a 20 μm photopatterned alignment cell. The filled cell was then cooled to the desired temperature (phase) using an Instec HCS 402 heat stage. The filled cells were cooled at 2°C/min to 70°C, then 0.5°C/min to 60°C, and 0.25°C/min at 0.5°C increments thereafter. Samples were equilibrated at the designated temperature for at least 2 minutes. Thereafter, the mixtures were photopolymerized for 10 minutes with 365 nm light at an intensity of 50 mW/cm$^2$. The polymer networks were extracted from the alignment cell and subsequently washed with acetone to remove any unreacted components.

*Materials characterization*

Phase transition temperatures and micrograph textures were obtained via polarized optical microscopy (POM) on a Nikon Eclipse Ci-POL in reflection mode with inline Instec HCS 402 heat stage. Kossel diagrams were captured on the same microscope with 100x oil-immersion objective (1.25na), Bertrand lens, and $\lambda_c$ = 440 nm x 10 nm bandpass filter.

Stress-strain measurements were performed on an RSA-G2 solids analyzer at a linear strain rate of 5% strain per minute. Glass transition temperatures were determined via differential scanning calorimetry (DSC) (TA Instruments Discovery DSC 2500). Transmission and reflection spectra of liquid crystal elastomers were collected with a Cary 7000 spectrometer (UV-Vis) utilizing a universal measurement accessory.

Gel fractions were obtained by recording the mass of the polymer sample before and after washing with dichloromethane for 24 hours.

## Acknowledgements

K.R.S. acknowledges fellowship support from the Department of Defense (DoD) through the National Defense Science & Engineering Graduate (NDSEG) Program. We thank Timothy Bunning for helpful conversations.

## Author Contributions

T.J.W. and K.R.S. designed the research. K.R.S. completed experiments and analyzed the data. K.R.S. and T.J.W. wrote the manuscript.

## Competing interests

The authors declare no competing financial interests.

Retention and Deformation of the Blue Phases in Liquid Crystalline Elastomers

Kyle R. Schlafmann[1], Timothy J. White[1,2,*]

# Supplementary Information

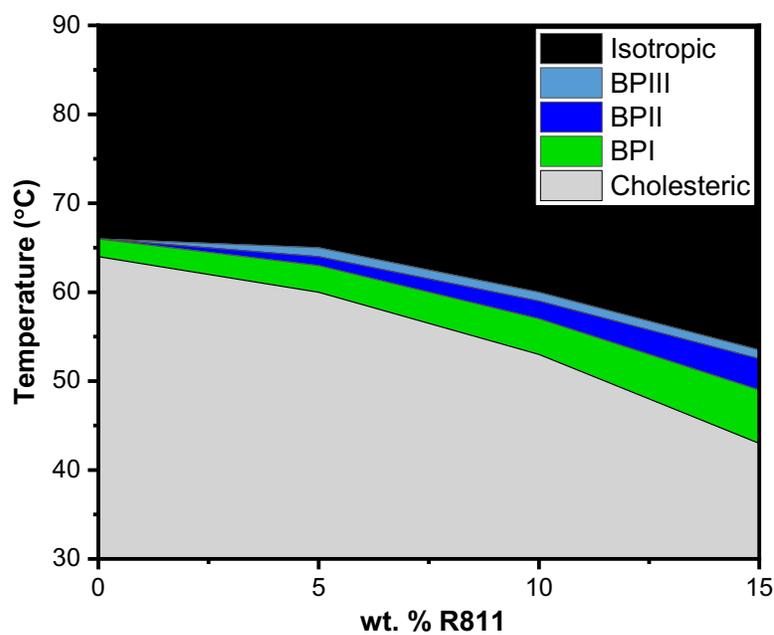

**Figure S1 | Blue phase stabilization**. Thermotropic phase windows for monomer mixtures (0.8:1 thiol to acrylate functional group ratio) with constant weighted average helical twisting power. The chiral diacrylate SLO 4151 has a helical twisting power of 8μm and the non-reactive chiral dopant R811 has a helical twisting power of 11μm).

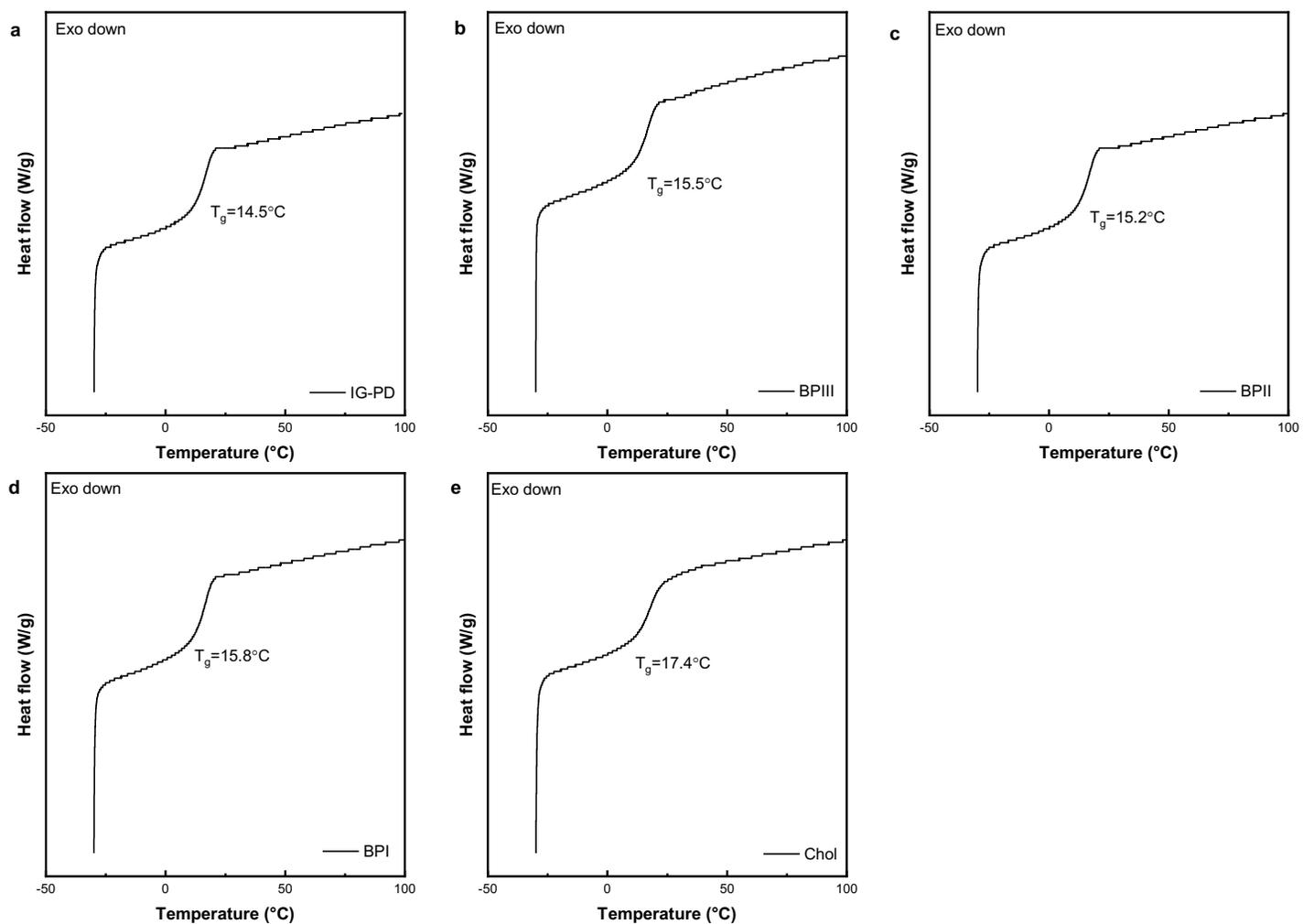

**Figure S2 | Glass transition temperatures of LCEs**. **a**, Isotropic genesis polydomain. **b**, Blue phase III. **c**, Blue phase II. **d**, Blue phase I. **e**, Cholesteric. Glass transition temperatures measured at the midpoint of the inflection point of the second heating curve (5°C/min) of a heat-cool-heat cycle.

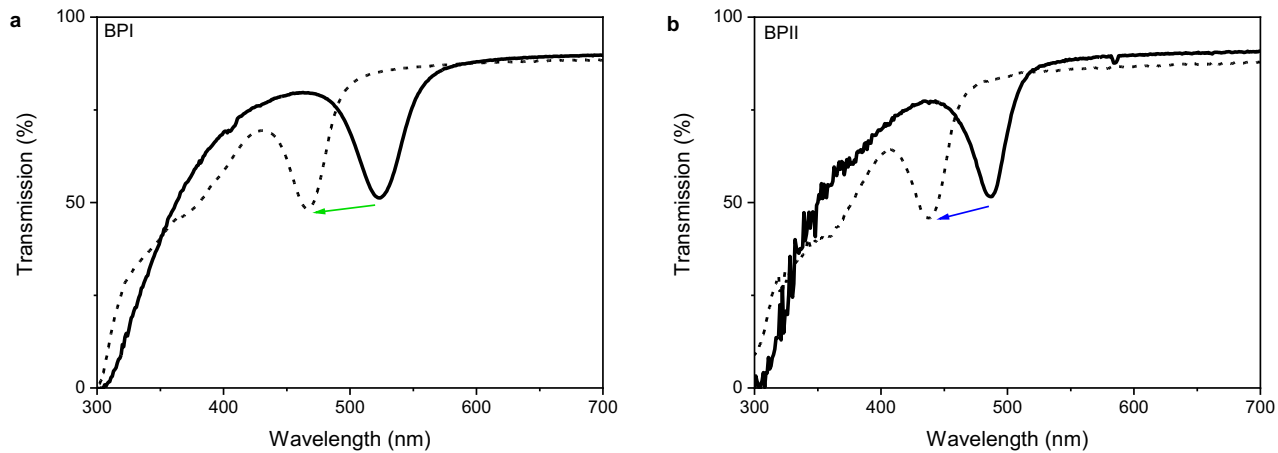

**Figure S3 | Wash effects. a**, UV-vis spectra of BPI LCE Before film extraction and solvent wash (solid line) and after (dashed line). **b,** BPII LCE (right).

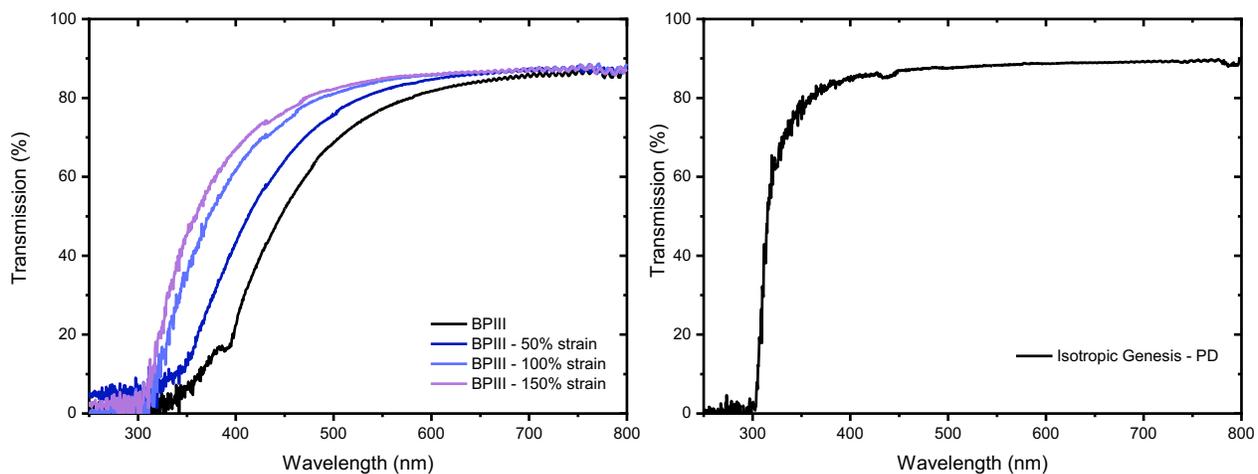

**Figure S4 | Spectra of amorphous phases**. **a**, UV-vis spectra of blue phase III (BPIII) LCE as a function of uniaxial strain. **b**, UV-vis spectra of isotropic genesis polydomain (IG-PD) LCE.

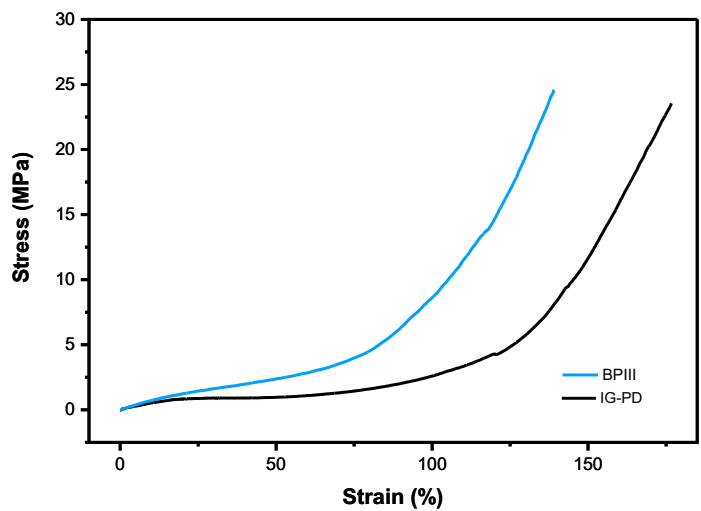

**Figure S5 | Stress-strain behavior of amorphous phase LCEs.**

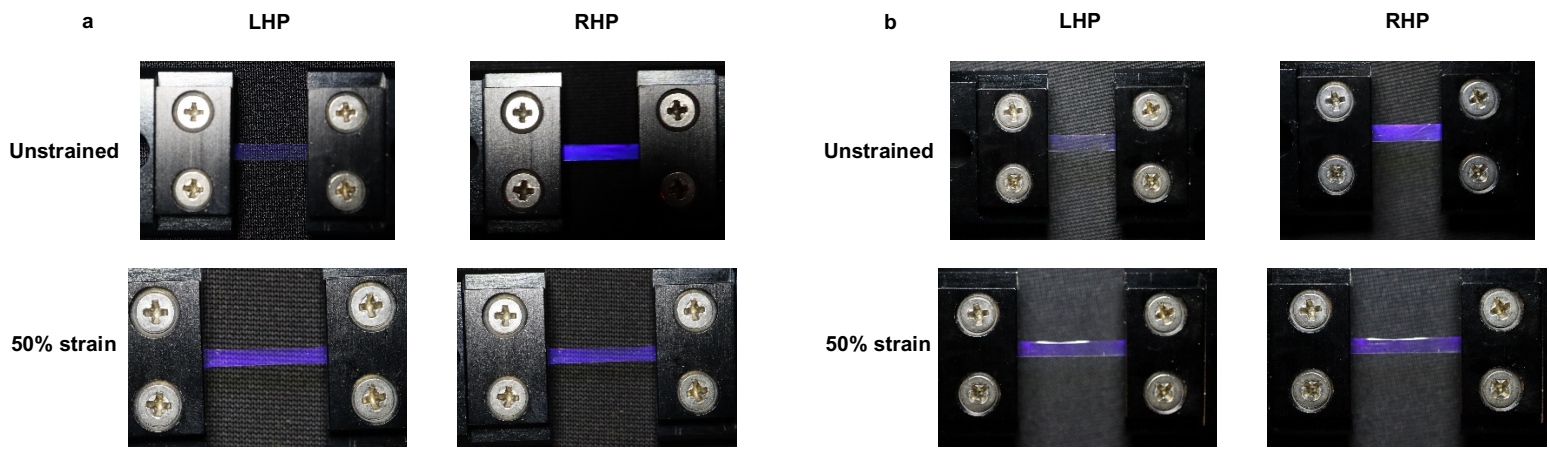

**Figure S6 | Reflection unpolarization. a**, Photographs of Blue Phase I (BPI) LCE illuminated by right-handed or left-handed circular polarization, with and without uniaxial deformation. **b**, Photographs of Blue Phase II (BPII) LCE illuminated by right-handed or left-handed circular polarization, with and without uniaxial deformation.

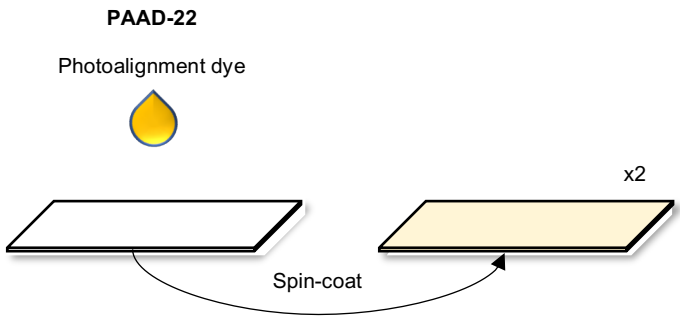
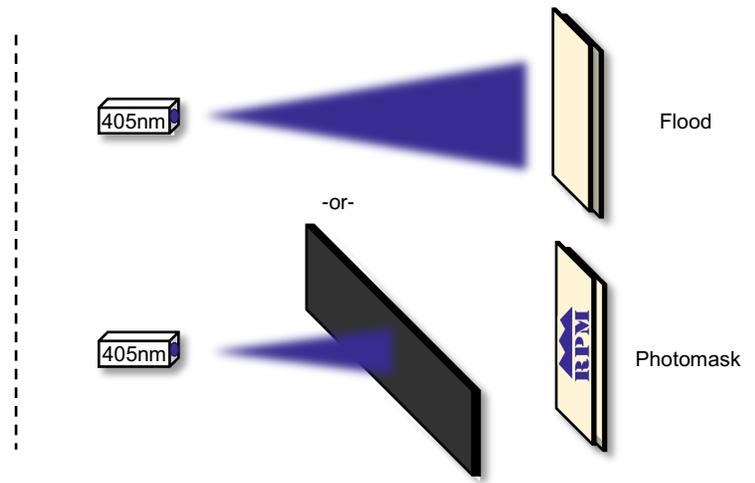

**Figure S7 | Photoalignment method**.